\documentclass[a4paper,11pt]{article}

\usepackage[dvips]{hyperref}
\hypersetup{
colorlinks=true,
linkcolor=magenta,
citecolor=blue,
urlcolor=black,
bookmarksnumbered=true,
bookmarkstype=toc}
\usepackage[margin=1in]{geometry}
\usepackage{amsmath,amsbsy,bm}
\usepackage[adobe-utopia,cal=cmcal]{mathdesign}
\usepackage[T1]{fontenc}
\usepackage{cite}
\usepackage{fancyhdr}
\usepackage[font=small,labelfont=bf]{caption}
\usepackage{subfigure}
\usepackage{tikz}
\usetikzlibrary{arrows}
\makeatletter

\@addtoreset{equation}{section}
\makeatother
\usepackage{titlesec}
\titleformat{\section}[block]{\filright\bfseries\mathversion{bold}}{\thesection.}{0.5em}{}[\titlerule]
\titleformat{\subsection}[block]{\filright\bfseries\mathversion{bold}}{\thesubsection.}{0.5em}{}
\usepackage{marvosym}
\makeatletter
\renewcommand*{\@fnsymbol}[1]{\ifcase#1\or \Letter\else\@ctrerr\fi}
\makeatother

\title{
\bf\large\mathversion{bold}Non-Abelian Monopole in the Parameter Space of Point-like Interactions
}
\author{
Satoshi Ohya${}^{\clubsuit,\heartsuit,}$\footnote{\tt\href{mailto:ohyasato@fjfi.cvut.cz}{ohyasato@fjfi.cvut.cz}}\\[2ex]
\textit{\small ${}^{\clubsuit}$Department of Physics, Faculty of Nuclear Sciences and Physical Engineering}\\
\textit{\small Czech Technical University in Prague}\\
\textit{\small Pohrani\v{c}n\'{i} 1288/1, 40501 D\v{e}\v{c}\'{i}n, Czech Republic}\\
\textit{\small ${}^{\heartsuit}$Doppler Institute for Mathematical Physics and Applied Mathematics}\\
\textit{\small Czech Technical University in Prague}\\
\textit{\small B\v{r}ehov\'{a} 7, 11519 Prague, Czech Republic}
}
\date{\small (Dated: \today)}

\begin{document}
\maketitle
\thispagestyle{fancy}
\renewcommand{\headrulewidth}{0pt}
\rhead{DI14-007}

\begin{abstract}
We study non-Abelian geometric phase in $\mathscr{N} = 2$ supersymmetric quantum mechanics for a free particle on a circle with two point-like interactions at antipodal points.
We show that non-Abelian Berry's connection is that of $SU(2)$ magnetic monopole discovered by Moody, Shapere and Wilczek in the context of adiabatic decoupling limit of diatomic molecule.
\end{abstract}

\tableofcontents

\newpage
\section{Introduction and summary} \label{sec:1}
\begin{figure}[t]
\center
\begin{tikzpicture}
\tikzstyle{every node}=[font=\scriptsize]
\begin{scope}
\draw (0,0) circle(1.5cm);
\filldraw[fill=white]
(-1.5,0) node[right]{$x=0$} node[left]{$U(2)$} circle(1.5pt)
(1.5,0) node[left]{$x=\ell$} node[right]{$U(2)$} circle(1.5pt);
\end{scope}
\begin{scope} [xshift=3cm]
\draw[->] (0,0) -- node[above]{folding} (1.5,0);
\end{scope}
\begin{scope} [xshift=6cm]
\def\L{3}
\draw (0,0) -- (\L,0);
\filldraw[fill=white]
(0,0) node[below]{$x=0$} node[above]{$U(2)$} circle(1.5pt)
(\L,0) node[below]{$x=\ell$} node[above]{$U(2)$} circle(1.5pt);
\end{scope}
\end{tikzpicture}
\caption{A circle with two point-like interactions at antipodal points.}
\label{fig:1}
\end{figure}
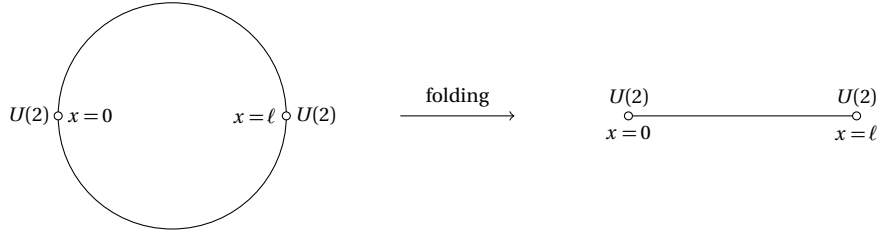

The purpose of this short note is to present a simple, easily solvable quantum mechanical model that exhibits doubly-degenerate energy spectrum and shows nontrivial non-Abelian geometric phase \cite{Wilczek:1984dh} under a cyclic adiabatic evolution of model parameters.
The keys to our model are supersymmetry and point-like interactions:
Supersymmetric quantum mechanics enjoys spectral degeneracy in general and hence provides a natural playground for studying non-Abelian geometric phase.
Point-like interactions, on the other hand, are known to be characterized by $U(2)$ family of boundary conditions in general \cite{Fulop:1999pf} and hence offers a simple yet nontrivial arena for studying geometric phase over the space of model parameters.
A simple example for such an interplay between point-like interactions and Abelian geometric phases has been elaborated in \cite{Cheon:2000tq}.
(It has also been briefly discussed in Ref.~\cite{Exner:1999}.)
There, it has been shown that Abelian Berry's connection is that of the $U(1)$ Dirac monopole in the subspace $U(2)/U(1)^{2} \cong SU(2)/U(1) \cong S^{2}$ of the group manifold $U(2)$.
In this note we present a kind of generalizations of \cite{Cheon:2000tq} to a non-Abelian situation.
For simplicity, we focus on quantum mechanics for a free spinless particle on a circle with two point-like interactions at antipodal points (see Figure \ref{fig:1}); that is, we consider a simple model in which the particle has no spin, the Hamiltonian is just minus the Laplacian, yet there are generic contact interactions consistent with probability current conservations (or, equivalently, unitarity of the theory).
The full parameter space of the model is $U(2) \times U(2)$, and energy levels are not degenerate in general.
However, there does exist a subspace $\mathcal{M}_{\text{SUSY}} \subset U(2) \times U(2)$ in which $\mathscr{N} = 2$ supersymmetry emerges and the energy spectrum exhibits two-fold degeneracy.
It should be noted that $\mathscr{N} = 2$ supersymmetry hidden behind degenerate spectra in exactly the same setup has been already studied in \cite{Nagasawa:2002un}.
In their analysis, however, the authors focused on positive semidefinite energy spectrum and overlooked the whole parameter space that exhibits two-fold degeneracy.
Though supersymmetry normally implies nonnegativity of energy eigenvalues, it turns out that it is too restrictive to focus on nonnegative spectrum in order to understand the entire parameter space of degenerate spectrum.
The point is a \textit{trivial} central extension of supersymmetry algebra, which is given by $\{Q_{\alpha}, Q_{\beta}\} = 2\delta_{\alpha\beta}(H + c)$ ($\alpha,\beta = 1,\cdots,\mathscr{N}$), where $Q_{\alpha}$ are self-adjoint supercharges, $H$ is a self-adjoint Hamiltonian and $c \in \mathbb{R}$ is a trivial center.
In any unitary representations of supersymmetry algebra, $H + c$ becomes a positive semidefinite operator such that spectrum of the Hamiltonian $\mathrm{Spec}\,(H)$ is bounded from below, $\mathrm{Spec}\,(H) \geq -c$.
This trivial central extension is really trivial -- because it can be removed by redefining the Hamiltonian, or shifting the origin of energy -- but it leads us to reveal the whole parameter space that enjoys spectral degeneracy in positive as well as negative energy spectra without spoiling supersymmetry.
In what follows we first show that the entire supersymmetric subspace $\mathcal{M}_{\text{SUSY}}$ of the model is given by the direct product:
\begin{align}
\mathcal{M}_{\text{SUSY}}
= 	U(1) \times U(2)/U(1)^{2}
\subset 	U(2) \times U(2). \label{eq:1.1}
\end{align}
In Ref.~\cite{Nagasawa:2002un} the first $U(1)$ factor was overlooked.
We will see that this $U(1)$ factor controls the energy spectrum (see Figure \ref{fig:2}) and enters the supersymmetry algebra through the trivial center $c$.
The second $U(2)/U(1)^{2}$ factor, on the other hand, does not appear in the energy spectrum at all -- but it \textit{does} appear in the energy eigenfunctions -- such that we obtain $\dim U(2)/U(1)^{2} = 2$ parameter family of isospectral systems of spectral multiplicity two.
Hence, as in the case of Abelian geometric phase in \cite{Cheon:2000tq}, the coset $U(2)/U(1)^{2} \cong S^{2}$ sets the scene for non-Abelian geometric phase.
The goal of this note is to compute non-Abelian Berry's connection over the space $\mathcal{M}_{\text{SUSY}}$ with fixed $U(1)$ parameter and show that an $SU(2)$ magnetic monopole appears in the 2-sphere $U(2)/U(1)^{2} \cong S^{2}$.

The rest of the note is organized as follows.
In section \ref{sec:2} we construct $\mathscr{N} = 2$ supersymmetric quantum mechanics on a circle of circumference $2\ell$ with two point-like interactions at $x = 0$ and $\ell$.
We explicitly derive the energy spectrum and energy eigenfunctions.
In section \ref{sec:3} we study non-Abelian geometric phase and show that non-Abelian Berry's connection on the coset $U(2)/U(1)^{2}$ is that of $SU(2)$ magnetic monopole found by Moody, Shapere and Wilczek in the context of adiabatic decoupling limit of ``slow'' and ``fast'' degrees of freedom in diatomic molecule \cite{Moody:1985ty}.
Computational details are relegated to appendix \ref{appendix:A}.

Throughout the note we will work in the units $\hbar = 2m = 1$, where $m$ is the mass of the particle.

\section{\texorpdfstring{$\mathscr{N}=2$}{N=2} mechanics} \label{sec:2}
$\mathscr{N} = 2$ supersymmetric quantum mechanics consists of four operators: a self-adjoint Hamiltonian $H$, two self-adjoint supercharges $Q_{1}$ and $Q_{2}$, and a fermion parity $(-1)^{F}$ which is a $\mathbb{Z}_{2}$-grading operator.
Usually, Hamiltonian is given from the beginning, but supercharges and fermion parity are not.
We have to find and construct $Q_{1}$, $Q_{2}$ and $(-1)^{F}$ in the course of solving a problem, which often requires a lot of guesswork.
In this section we shall construct these operators explicitly and solve the energy eigenvalue problems.

We note that most of the materials presented in this section are already known and discussed in Ref.~\cite{Nagasawa:2002un} (see also \cite{Ohya:2012qz}).
But some of the materials are new.
The main new result is that, as mentioned in the previous section, the whole parameter space $\mathcal{M}_{\text{SUSY}} \subset U(2) \times U(2)$ that admits $\mathscr{N} = 2$ supersymmetry is given by the direct product \eqref{eq:1.1}.
In order to make this note self-contained, we will reproduce all the relevant results.
We also note that this section is mostly based on our previous work \cite{Ohya:2012qz}, with some minor changes to make things clearer.

\subsection{\texorpdfstring{$\mathbb{Z}_{2}$}{Z2}-graded Hilbert space} \label{sec:2.1}
To begin with, let us first fix the notations.
Let $\psi(x)$ ($0 < x < 2\ell$) be a wavefunction on the circle of circumference $2\ell$.
For the following discussions it is convenient to consider the wavefunction on the upper- and lower-semicircles separately, $\psi(x)$ for $0 < x < \ell$ and $\psi(x)$ for $\ell < x < 2\ell$, and embed them into a single 2-component vector-valued function $\bm{\psi}$ on the interval $(0, \ell)$ as follows:
\begin{align}
\bm{\psi}(x)
&= 	\begin{pmatrix}
	\psi(x) \\
	\psi(2\ell - x)
	\end{pmatrix},
	\quad 0 < x < \ell. \label{eq:2.1}
\end{align}
Technically, we consider the following Hilbert space $\mathcal{H}$:
\begin{align}
\mathcal{H}
&= 	L^{2}(0, \ell) \oplus L^{2}(\ell, 2\ell)\nonumber\\
&\cong L^{2}(0, \ell) \otimes \mathbb{C}^{2}, \label{eq:2.2}
\end{align}
where $L^{2}(0, \ell)$ is a set of square integrable functions on the interval $(0, \ell)$.
In this way quantum mechanics on a circle is always mapped into quantum mechanics on an interval with vector-valued wavefunction, which is often called the folding trick; see Figure \ref{fig:1}.
Point-like interactions at antipodal points on the circle are then translated into the problem of boundary conditions at the boundaries of the interval.

Let us next consider the Hamiltonian operator for a free particle on the interval.
The free Hamiltonian $H$ acting on $\mathcal{H} \ni \bm{\psi}$ is given by the $2 \times 2$ diagonal matrix
\begin{align}
H
= 	\begin{pmatrix}
	h	& 0 \\
	0 	& h
	\end{pmatrix}
= 	h \otimes \bm{1}_{2} \quad\text{with}\quad h = -\frac{d^{2}}{dx^{2}}. \label{eq:2.3}
\end{align}
Given an observable, we have to specify the domain in which the observable becomes a self-adjoint operator.
For the case of Hamiltonian, this is physically equivalent to the requirement of probability current conservations at the boundaries, $j(0) = 0 = j(\ell)$, where $j(x) = -i(\bm{\psi}^{\dagger}(x)\bm{\psi}^{\prime}(x) - \bm{\psi}^{\prime\dagger}(x)\bm{\psi}(x))$.
(Prime (${}^{\prime}$) indicates the derivative with respect to $x$.)
As is well-known \cite{Fulop:1999pf}, this requirement leads to the following $U(2) \times U(2)$ family of boundary conditions:
\begin{subequations}
\begin{align}
(\bm{1}_{2} - U)\bm{\psi}(0) - iL_{0}(\bm{1}_{2} + U)\bm{\psi}^{\prime}(0)
&= 	\bm{0}, \quad U \in U(2), \label{eq:2.4a}\\
(\bm{1}_{2} - \Bar{U})\bm{\psi}(\ell) + iL_{0}(\bm{1}_{2} + \Bar{U})\bm{\psi}^{\prime}(\ell)
&= 	\bm{0}, \quad \Bar{U} \in U(2), \label{eq:2.4b}
\end{align}
\end{subequations}
where $\bm{1}_{2}$ stands for the $2 \times 2$ unit matrix.
$L_{0}$ is an arbitrary scale parameter of length dimension one which must be introduced to adjust the scaling dimension of the first and second terms of the boundary conditions.
We note that without any loss of generality $L_{0}$ can be taken to be positive.
Eqs.~\eqref{eq:2.4a} and \eqref{eq:2.4b} describe the most general point-like interactions in unitary quantum mechanics.

Let us next move on to the construction of fermion parity $(-1)^{F}$ on $\mathcal{H}$.
Since the fermion parity should satisfy $((-1)^{F})^{2} = 1$ and have eigenvalues $+1$ and $-1$, we have to find a $\mathbb{Z}_{2}$ operator whose square becomes the identity.
A naive guess is $(-1)^{F} = \mathcal{P}$, where $\mathcal{P}: x \mapsto 2\ell - x$ is the parity operator on the circle and acts on the vector wavefunction as $(\mathcal{P}\bm{\psi})(x) = \bm{\psi}(2\ell - x) = \sigma_{1}\bm{\psi}(x)$.\footnote{Indeed, in many examples of hidden $\mathscr{N} = 2$ supersymmetry, parity provides the $\mathbb{Z}_{2}$-grading operator; see e.g. Ref.~\cite{Jakubsky:2010ki}.}
However, we do not need to be stuck in this one particular discrete transformation.
As discussed in \cite{Nagasawa:2002un}, the fermion parity is generally given by the following unitary transformation on $\mathcal{H}$:\footnote{In Ref.~\cite{Nagasawa:2002un}, the $\mathbb{Z}_{2}$ transformation \eqref{eq:2.5} is denoted by $\mathcal{G}$ and defined as $\mathcal{G} = \Vec{\alpha} \cdot \Vec{\mathcal{P}}$, where $\Vec{\alpha}$ is a real unit 3-vector and $\Vec{\mathcal{P}} = (\mathcal{P}_{1}, \mathcal{P}_{2}, \mathcal{P}_{3})$ is a vector of discrete transformations given by $\mathcal{P}_{1} = \mathcal{P}$ (parity), $\mathcal{P}_{3} = \mathcal{R}$ (half-reflection) and $\mathcal{P}_{2} = -i\mathcal{R}\mathcal{P}$. These discrete transformations act on the vector-valued wavefunction as $(\mathcal{P}_{1}\bm{\psi})(x) = \sigma_{1}\bm{\psi}(x)$, $(\mathcal{P}_{2}\bm{\psi})(x) = \sigma_{2}\bm{\psi}(x)$ and $(\mathcal{P}_{3}\bm{\psi})(x) = \sigma_{3}\bm{\psi}(x)$ such that the definition $\mathcal{G} = \Vec{\alpha} \cdot \Vec{P}$ is equivalent to the parameterization \eqref{eq:2.12}.}
\begin{align}
\mathcal{Z}: \bm{\psi}(x) \mapsto (\mathcal{Z}\bm{\psi})(x) := Z\bm{\psi}(x), \label{eq:2.5}
\end{align}
where $Z \in U(2)$ is a generic $2 \times 2$ traceless hermitian unitary matrix that satisfies $Z^{2} = \bm{1}_{2}$ (i.e. $Z = Z^{\dagger} = Z^{-1}$) and $Z \neq \pm \bm{1}_{2}$.
Obviously thus defined unitary transformation satisfies $\mathcal{Z}^{2} = 1$.
Notice that $Z$ has the eigenvalues $+1$ and $-1$ such that it can be written as the following spectral decomposition:
\begin{align}
Z = (+1)P_{+} + (-1)P_{-}, \label{eq:2.6}
\end{align}
where
\begin{align}
P_{\pm} = \frac{\bm{1}_{2} \pm Z}{2} \label{eq:2.7}
\end{align}
are hermitian projection operators that fulfil the orthonormalities $P_{\alpha}P_{\beta} = \delta_{\alpha\beta}P_{\beta}$ ($\alpha, \beta = \pm$), the completeness $P_{+} + P_{-} = \bm{1}_{2}$ and the hermiticity $P_{\pm}^{\dagger} = P_{\pm}$.

Now we have found the fermion parity $(-1)^{F} = \mathcal{Z}$, but this is not the end of the story because the fermion parity must be a symmetry of the system.
In general, a unitary transformation on $\mathcal{H}$ is said to be a symmetry if it commutes with the Hamiltonian and further preserves the self-adjoint domain of the Hamiltonian (i.e., leaves the boundary conditions unchanged) \cite{Fulop:2003}.
In our problem, the first condition is obviously satisfied because the Hamiltonian $H = h \otimes \bm{1}_{2}$ and the fermion parity $\mathcal{Z} = 1 \otimes Z$ trivially commutes, $[H, \mathcal{Z}] = 0$.
The second condition is, however, nontrivial.
The unitary transformation \eqref{eq:2.5} leaves the boundary conditions \eqref{eq:2.4a} and \eqref{eq:2.4b} unchanged if and only if the unitary matrices $Z$, $U$ and $\Bar{U}$ are all simultaneously diagonalizable and satisfy the conditions $[U, Z] = [\Bar{U}, Z] = 0$.\footnote{\label{footnote:3}Here is the proof \cite{Fulop:2003}. Let $\bm{\psi} \in \mathcal{H}$ satisfy the boundary conditions \eqref{eq:2.4a} and \eqref{eq:2.4b}. Then the transformed state $(\mathcal{Z}\bm{\psi})(x) = Z\bm{\psi}(x)$ satisfies the following boundary conditions:
\begin{align}
(\bm{1}_{2} - ZUZ)(\mathcal{Z}\bm{\psi})(0) - iL_{0}(\bm{1}_{2} + ZUZ)(\mathcal{Z}\bm{\psi})^{\prime}(0)
= 	\bm{0}
\quad\text{and}\quad
(\bm{1}_{2} - Z\Bar{U}Z)(\mathcal{Z}\bm{\psi})(\ell) + iL_{0}(\bm{1}_{2} + Z\Bar{U}Z)(\mathcal{Z}\bm{\psi})^{\prime}(\ell)
= 	\bm{0}, \nonumber
\end{align}
from which we deduce that $\mathcal{Z}$ induces the maps $U \stackrel{\mathcal{Z}}{\mapsto} ZUZ$ and $\Bar{U} \stackrel{\mathcal{Z}}{\mapsto} Z\Bar{U}Z$.
If $U = ZUZ$ and $\Bar{U} = Z\Bar{U}Z$, the boundary conditions remain unchanged and $\mathcal{Z}$ becomes the symmetry of the system.}
It is easy to see that the general solutions to the conditions $[U, Z] = [\Bar{U}, Z] = 0$ are parameterized as follows:
\begin{subequations}
\begin{align}
U
&= 	\mathrm{e}^{i\alpha_{+}}P_{+} + \mathrm{e}^{i\alpha_{-}}P_{-}, \label{eq:2.8a}\\
\Bar{U}
&= 	\mathrm{e}^{i\Bar{\alpha}_{+}}P_{+} + \mathrm{e}^{i\Bar{\alpha}_{-}}P_{-}, \label{eq:2.8b}
\end{align}
\end{subequations}
where $\alpha_{\pm}$ and $\Bar{\alpha}_{\pm}$ are eigenphases of $U$ and $\Bar{U}$, respectively.
The fermion parity $(-1)^{F} = \mathcal{Z}$ becomes the symmetry if and only if the boundary conditions are specified by these unitary matrices $U$ and $\Bar{U}$.

This $\mathcal{Z}$-invariance extremely simplifies the model.
First of all, the Hilbert space splits into two orthogonal subspaces and can be written as the following direct sum:
\begin{align}
\mathcal{H} = \mathcal{H}_{+} \oplus \mathcal{H}_{-}, \label{eq:2.9}
\end{align}
where $\mathcal{H}_{\pm} = \{\bm{\psi}_{\pm} \in \mathcal{H} \mid Z\bm{\psi}_{\pm} = \pm\bm{\psi}_{\pm}\}$ are ``bosonic'' and ``fermionic'' sectors of the model, respectively.
Second of all, by letting $\bm{e}_{+}$ and $\bm{e}_{-}$ be orthonormal eigenvectors of $Z$ satisfying $Z\bm{e}_{\pm} = \pm\bm{e}_{\pm}$, $\bm{e}_{\alpha}^{\dagger}\bm{e}_{\beta} = \delta_{\alpha\beta}$ ($\alpha, \beta = \pm$) and $\bm{e}_{\pm}\bm{e}_{\pm}^{\dagger} = P_{\pm}$, one immediately sees that the wavefunction $\bm{\psi}_{\pm} \in \mathcal{H}_{\pm}$ can be written as follows:
\begin{align}
\bm{\psi}_{\pm}(x) = \psi_{\pm}(x)\bm{e}_{\pm}, \label{eq:2.10}
\end{align}
where $\psi_{\pm}(x) = \bm{e}_{\pm}^{\dagger}\bm{\psi}(x)$.
In what follows we call the set of orthonormal eigenvectors $\{\bm{e}_{+}, \bm{e}_{-}\}$ the basis and $\psi_{\pm}$ the components.
It is easy to check that, in terms of the components, the boundary conditions \eqref{eq:2.4a} and \eqref{eq:2.4b} with the unitary matrices \eqref{eq:2.8a} and \eqref{eq:2.8b} boil down to the following four independent Robin boundary conditions:
\begin{subequations}
\begin{align}
\sin\left(\frac{\alpha_{\pm}}{2}\right)\psi_{\pm}(0) + L_{0}\cos\left(\frac{\alpha_{\pm}}{2}\right)\psi_{\pm}^{\prime}(0)
&= 	0, \label{eq:2.11a}\\
\sin\left(\frac{\Bar{\alpha}_{\pm}}{2}\right)\psi_{\pm}(\ell) - L_{0}\cos\left(\frac{\Bar{\alpha}_{\pm}}{2}\right)\psi_{\pm}^{\prime}(\ell)
&= 	0. \label{eq:2.11b}
\end{align}
\end{subequations}
Hence, in the diagonal basis $\{\bm{e}_{+}, \bm{e}_{-}\}$, the system is reduced to two separate systems on the interval $(0, \ell)$ described by the Schr\"{o}dinger equation $h\psi_{\pm} = E\psi_{\pm}$ with the boundary conditions \eqref{eq:2.11a} and \eqref{eq:2.11b}.

We would like to add a few comments before going into the construction of supercharges.
\begin{itemize}
\item \textit{$\mathbb{Z}_{2}$-symmetric parameter space.}
The parameter space of this $\mathbb{Z}_{2}$-symmetric theory is given by the direct product of 4-torus and 2-sphere with an identification of points under the action of the discrete group $\mathbb{Z}_{2}$.
To see this, let us first consider the diagonalization of $Z$.
Any hermitian traceless unitary matrix $Z$ can be diagonalized as $Z = V\left(\begin{smallmatrix}+1&0\\0&-1\end{smallmatrix}\right)V^{\dagger}$ by some $2 \times 2$ unitary matrix $V \in U(2)$.
Note, however, that since $Z$ is invariant under the right multiplication $V \mapsto V\left(\begin{smallmatrix}\mathrm{e}^{i\beta_{1}}&0\\0&\mathrm{e}^{i\beta_{2}}\end{smallmatrix}\right)$, where $\mathrm{e}^{i\beta_{1}}$ and $\mathrm{e}^{i\beta_{2}}$ are arbitrary $U(1)$ phases, $Z$ should be regarded as an element of the coset $U(2)/U(1)^{2}$, which is the complete flag manifold and isomorphic to 2-sphere $S^{2}$.
It can also be seen by explicit parameterization that the parameter space of $Z$ is $S^{2}$.
As is well-known, any $2 \times 2$ traceless hermitian unitary matrix $Z$ can be written as
\begin{align}
Z = \bm{n} \cdot \bm{\sigma}, \label{eq:2.12}
\end{align}
where $\bm{\sigma} = (\sigma_{1}, \sigma_{2}, \sigma_{3})$ is a vector of the Pauli matrices and $\bm{n} = (n_{1}, n_{2}, n_{3})$ is a real unit vector fulfilling the condition $n_{1}^{2} + n_{2}^{2} + n_{3}^{2} = 1$; that is, $\bm{n}$ parameterizes the unit 2-sphere $S^{2}$.
Along with this parameter $\bm{n}$, we have four distinct parameters $\{\alpha_{\pm}, \Bar{\alpha}_{\pm}\}$ that parameterize the eigenvalues of $U$ and $\Bar{U}$ (i.e. the maximal torus $U(1)^{4} \cong T^{4}$ in $U(2) \times U(2)$).
Hence one may write $U = U(\alpha_{+}, \alpha_{-}; \bm{n})$ and $\Bar{U} = \Bar{U}(\Bar{\alpha}_{+}, \Bar{\alpha}_{-}; \bm{n})$.
It should be noted here that $U$ and $\Bar{U}$ in Eqs.~\eqref{eq:2.8a} and \eqref{eq:2.8b} with the parameterization $P_{\pm} = (\bm{1}_{2} \pm \bm{n}\cdot\bm{\sigma})/2$ satisfy the identities
\begin{align}
U(\alpha_{+}, \alpha_{-}; \bm{n})
= 	U(\alpha_{-}, \alpha_{+}; -\bm{n})
\quad\text{and}\quad
\Bar{U}(\Bar{\alpha}_{+}, \Bar{\alpha}_{-}; \bm{n})
= 	\Bar{U}(\Bar{\alpha}_{-}, \Bar{\alpha}_{+}; -\bm{n}). \label{eq:2.13}
\end{align}
Hence the $\mathbb{Z}_{2}$-symmetric parameter space $\mathcal{M}$ is given by the direct product of $U(1)^{4} \cong T^{4}$ and $U(2)/U(1)^{2} \cong S^{2}$ with the $\mathbb{Z}_{2}$ identification $(\alpha_{+}, \alpha_{-}; \Bar{\alpha}_{+}, \Bar{\alpha}_{-}; \bm{n}) \sim (\alpha_{-}, \alpha_{+}; \Bar{\alpha}_{-}, \Bar{\alpha}_{+}; -\bm{n})$:
\begin{align}
\mathcal{M}
= 	\frac{U(1)^{4} \times U(2)/U(1)^{2}}{\mathbb{Z}_{2}}
\cong \frac{T^{4} \times S^{2}}{\mathbb{Z}_{2}}. \label{eq:2.14}
\end{align}

\item \textit{Weyl pair $(\mathcal{X}, \mathcal{Z})$.}
Once given the $\mathbb{Z}_{2}$ transformation $\mathcal{Z}$, we can introduce another $\mathbb{Z}_{2}$ unitary transformation $\mathcal{X}$ on $\mathcal{H}$ that satisfies the following anticommutation relation:
\begin{align}
\mathcal{X}\mathcal{Z} = -\mathcal{Z}\mathcal{X}. \label{eq:2.15}
\end{align}
Indeed, such unitary operator $\mathcal{X}$ can be defined as follows:
\begin{align}
\mathcal{X}: \bm{\psi}(x) \mapsto (\mathcal{X}\bm{\psi})(x) := X\bm{\psi}(x), \label{eq:2.16}
\end{align}
where $X$ is a $2 \times 2$ traceless hermitian unitary matrix satisfying $X = X^{\dagger} = X^{-1}$ and given by $X = \bm{e}_{+}\bm{e}_{-}^{\dagger} + \bm{e}_{-}\bm{e}_{+}^{\dagger}$.
By construction, the unitary matrices $X = \bm{e}_{+}\bm{e}_{-}^{\dagger} + \bm{e}_{-}\bm{e}_{+}^{\dagger}$ and $Z = \bm{e}_{+}\bm{e}_{+}^{\dagger} - \bm{e}_{-}\bm{e}_{-}^{\dagger}$ obviously anticommute with each other, $XZ = -ZX$, so do the operators $\mathcal{X}$ and $\mathcal{Z}$.
Operators that satisfy the relation \eqref{eq:2.15} are called a Weyl pair $(\mathcal{X}, \mathcal{Z})$ \cite{Kibler:2009} and essential ingredients of $\mathscr{N} = 2$ supersymmetric quantum mechanics.
The point is that the above constructed unitary matrix $X$ maps the eigenvectors $\bm{e}_{\pm}$ to $\bm{e}_{\mp}$:
\begin{align}
X\bm{e}_{\pm} = \bm{e}_{\mp}. \label{eq:2.17}
\end{align}
We will see in the next section that $X$ provides a building block to construct nilpotent supercharges $Q^{\pm}: \mathcal{H}_{\pm} \to \mathcal{H}_{\mp}$ in a basis independent fashion.
(Nilpotent supercharges $Q^{\pm}$ and self-adjoint supercharges $Q_{1,2}$ are related as $Q^{\pm} = (Q_{1} \pm iQ_{2})/2$.)
Another important property of $\mathcal{X}$ is its aspect of spectrum-preserving transformation.
To see this, suppose that the system is invariant under $\mathcal{Z}$ and the boundary conditions are specified by the unitary matrices \eqref{eq:2.8a} and \eqref{eq:2.8b}.
Then, it is easy to see that the unitary transformation \eqref{eq:2.16} induces the following maps (cf. footnote \ref{footnote:3}):
\begin{subequations}
\begin{align}
U(\alpha_{+}, \alpha_{-}; \bm{n})
&\stackrel{\mathcal{X}}{\mapsto}
	XU(\alpha_{+}, \alpha_{-}; \bm{n})X
= 	U(\alpha_{-}, \alpha_{+}; \bm{n}), \label{eq:2.18a}\\
\Bar{U}(\Bar{\alpha}_{+}, \Bar{\alpha}_{-}; \bm{n})
&\stackrel{\mathcal{X}}{\mapsto}
	X\Bar{U}(\Bar{\alpha}_{+}, \Bar{\alpha}_{-}; \bm{n})X
= 	\Bar{U}(\Bar{\alpha}_{-}, \Bar{\alpha}_{+}; \bm{n}), \label{eq:2.18b}
\end{align}
which follow from the identities $XP_{\pm}X = P_{\mp}$.
Since $\mathcal{X} = 1 \otimes X$ trivially commutes with $H = h \otimes \bm{1}_{2}$ and hence preserves the energy spectrum, two different systems specified by the unitary matrices $\{U(\alpha_{+}, \alpha_{-}; \bm{n}), \Bar{U}(\Bar{\alpha}_{+}, \Bar{\alpha}_{-}; \bm{n})\}$ and $\{U(\alpha_{-}, \alpha_{+}; \bm{n}), \Bar{U}(\Bar{\alpha}_{-}, \Bar{\alpha}_{+}; \bm{n})\}$ are isospectral.
We will see in the next section that $\mathcal{X}$ also provides a building block to construct a spectral duality between two different supersymmetric quantum mechanics.
\end{subequations}
\end{itemize}

\subsection{Supersymmetric boundary conditions} \label{sec:2.2}
Now let us move on to the construction of supercharges $Q_{\alpha}$.
Without a trivial central extension, supercharges and Hamiltonian are related as $H = (Q_{\alpha})^{2}$ such that $Q_{\alpha}$ are given by square roots of the Hamiltonian.
Since the free Hamiltonian is just the square of momentum operator $p = -id/dx$, in the absence of the trivial center, $Q_{\alpha}$ are basically given by the momentum operator.
With a trivial central extension $c$, however, $Q_{\alpha}$ and $H$ are related as $H = (Q_{\alpha})^{2} - c$ so that we are allowed to factorize the Hamiltonian up to a constant shift $c$.
The trivial center $c$ depends on the model of course, and we have to find it in the course of solving the problem.
To do this, the following two observations are crucial.
The first is that, except for $\alpha_{\pm} = \pi$ and $\Bar{\alpha}_{\pm} = \pi$ $\pmod{2\pi}$, at which the boundary conditions become the Dirichlet boundary conditions, the $\mathcal{Z}$-invariant boundary conditions \eqref{eq:2.11a} and \eqref{eq:2.11b} can be written as follows:
\begin{subequations}
\begin{align}
+ \psi_{\pm}^{\prime}(0) + \frac{1}{L(\alpha_{\pm})}\psi_{\pm}(0)
&= 	0, \label{eq:2.19a}\\
- \psi_{\pm}^{\prime}(\ell) + \frac{1}{L(\Bar{\alpha}_{\pm})}\psi_{\pm}(\ell)
&= 	0, \label{eq:2.19b}
\end{align}
\end{subequations}
where $L(\alpha) := L_{0}\cot(\alpha/2)$.
This motivates us to introduce the following one-parameter family of first-order differential operators:
\begin{align}
A^{\pm}_{\alpha}
&= 	\pm\frac{d}{dx} + \frac{1}{L(\alpha)}, \label{eq:2.20}
\end{align}
in terms of which the equations \eqref{eq:2.19a} and \eqref{eq:2.19b} are written as $(A^{+}_{\alpha_{\pm}}\psi_{\pm})(0) = (A^{-}_{-\alpha_{\pm}}\psi_{\pm})(0) = 0$ and $(A^{-}_{\alpha_{\pm}}\psi_{\pm})(\ell) = (A^{+}_{-\alpha_{\pm}}\psi_{\pm})(\ell) = 0$.
(Note that $A^{+}_{\alpha} = -A^{-}_{-\alpha}$.)
The second is that, by using these first-order differential operators, the free Hamiltonian $h = -d^{2}/dx^{2}$ is factorized up to a constant shift:
\begin{align}
h
&= 	A^{-}_{\alpha}A^{+}_{\alpha} - \frac{1}{L(\alpha)^{2}} \nonumber\\
&= 	A^{+}_{\alpha}A^{-}_{\alpha} - \frac{1}{L(\alpha)^{2}}. \label{eq:2.21}
\end{align}
The Schr\"{o}dinger equations $h\psi_{\pm} = E\psi_{\pm}$ can then be written into the following forms:
\begin{subequations}
\begin{align}
A^{-}_{\alpha}A^{+}_{\alpha}\psi_{+}(x)
&= 	\left[E + \frac{1}{L(\alpha)^{2}}\right]\psi_{+}(x), \label{eq:2.22a}\\
A^{+}_{\alpha}A^{-}_{\alpha}\psi_{-}(x)
&= 	\left[E + \frac{1}{L(\alpha)^{2}}\right]\psi_{-}(x), \label{eq:2.22b}
\end{align}
\end{subequations}
which imply the following supersymmetry relations:
\begin{subequations}
\begin{align}
A^{+}_{\alpha}\psi_{+}(x)
&= 	\sqrt{E + \frac{1}{L(\alpha)^{2}}}\psi_{-}(x), \label{eq:2.23a}\\
A^{-}_{\alpha}\psi_{-}(x)
&= 	\sqrt{E + \frac{1}{L(\alpha)^{2}}}\psi_{+}(x). \label{eq:2.23b}
\end{align}
\end{subequations}
We have almost done and are ready to construct the supercharges, but before doing that, we first classify the boundary conditions invariant under the supersymmetry transformations \eqref{eq:2.23a} and \eqref{eq:2.23b}.
Boundary conditions are said to be supersymmetric if the transformed state $A^{+}_{\alpha}\psi_{+}$ ($A^{-}_{\alpha}\psi_{-}$) satisfies the boundary condition for $\psi_{-}$ ($\psi_{+}$).
If we impose the Robin boundary condition at $x = 0$ on the ``bosonic'' sector, $(A^{+}_{\alpha}\psi_{+})(0) = 0$, which corresponds to the choice $\alpha_{+} = \alpha \neq \pi \pmod{2\pi}$, the supersymmetry relation \eqref{eq:2.23a} says that the ``fermionic'' sector should satisfy the Dirichlet boundary condition, $\psi_{-}(0) = 0$, which corresponds to the choice $\alpha_{-} = \pi \pmod{2\pi}$.
On the other hand, if we impose the Dirichlet boundary condition at $x = 0$ on the ``bosonic'' sector, $\psi_{+}(0) = 0$, which corresponds to the choice $\alpha_{+} = \pi \pmod{2\pi}$, the supersymmetry relation \eqref{eq:2.23b} says that the ``fermionic'' sector should satisfy the Robin boundary condition, $(A^{-}_{\alpha}\psi_{-})(0) = 0$, which corresponds to the choice $\alpha_{-} = -\alpha \neq \pi \pmod{2\pi}$.
Obviously similar arguments hold true for the boundary conditions at $x = \ell$.
Hence there are two distinct possibilities at each boundary, resulting in the following $2 \times 2 = 4$ distinct supersymmetric boundary conditions:
\begin{subequations}
\begin{alignat}{3}
&\text{type (D, D):}&~
&\begin{cases}
\psi_{+}(0) = 0 = (A_{\alpha}^{-}\psi_{-})(0), \\
\psi_{+}(\ell) = 0 = (A_{\alpha}^{-}\psi_{-})(\ell),
\end{cases}& \label{eq:2.24a}\\
&\text{type (R, R):}&
&\begin{cases}
(A_{\alpha}^{+}\psi_{+})(0) = 0 = \psi_{-}(0), \\
(A_{\alpha}^{+}\psi_{+})(\ell) = 0 = \psi_{-}(\ell),
\end{cases}& \label{eq:2.24b}\\
&\text{type (D, R):}&
&\begin{cases}
\psi_{+}(0) = 0 = (A_{\alpha}^{-}\psi_{-})(0), \\
(A_{\alpha}^{+}\psi_{+})(\ell) = 0 = \psi_{-}(\ell),
\end{cases}& \label{eq:2.24c}\\
&\text{type (R, D):}&
&\begin{cases}
(A_{\alpha}^{+}\psi_{+})(0) = 0 = \psi_{-}(0), \\
\psi_{+}(\ell) = 0 = (A_{\alpha}^{-}\psi_{-})(\ell),
\end{cases}& \label{eq:2.24d}
\end{alignat}
\end{subequations}
where ``D'' and ``R'' stand for Dirichlet and Robin, respectively, and (D, R), for example, means that the ``boson'' state $\bm{\psi}_{+} \in \mathcal{H}_{+}$ satisfies the Dirichlet boundary condition at $x = 0$ and Robin boundary condition at $x = \ell$, and so on.
It is easy to check that the type (D, D), (R, R), (D, R) and (R, D) boundary conditions correspond to the parameter choices $(\alpha_{+}, \alpha_{-}; \Bar{\alpha}_{+}, \Bar{\alpha}_{-}) = (\pi, -\alpha; \pi, \alpha)$, $(\alpha, \pi; -\alpha, \pi)$, $(\pi, -\alpha; -\alpha, \pi)$ and $(\alpha, \pi; \pi, \alpha)$ $\pmod{2\pi}$, respectively; see Table \ref{tab:1}.
Irrespective of these four types, the model parameters we have are $\alpha$ and $\bm{n}$, implying that the parameter space of supersymmetric boundary conditions is given by the direct product
\begin{align}
\mathcal{M}_{\text{SUSY}}
= 	U(1) \times U(2)/U(1)^{2}
\cong S^{1} \times S^{2}. \label{eq:2.25}
\end{align}
This is the main result of this section.

\begin{table}[t]
\begin{center}
\begin{tabular}{c|ccc}
\hline
boundary conditions 	& $U$ \& $\Bar{U}$ 						& Witten index 	& supersymmetry \\
\hline
type (D, D) 		& $U = U(\pi, -\alpha; \bm{n})$ 				& $-1$ 		& unbroken \\
				& $\Bar{U} = \Bar{U}(\pi, +\alpha; \bm{n})$ 	& 			& \\
type (R, R) 		& $U = U(+\alpha, \pi; \bm{n})$ 				& $+1$ 		& unbroken \\
				& $\Bar{U} = \Bar{U}(-\alpha, \pi; \bm{n})$ 	& 			& \\
type (D, R) 		& $U = U(\pi, -\alpha; \bm{n})$ 				& $0$ 		& spontaneously broken \\
				& $\Bar{U} = \Bar{U}(-\alpha, \pi; \bm{n})$ 	& 			& \\
type (R, D) 		& $U = U(\alpha, \pi; \bm{n})$ 				& $0$ 		& spontaneously broken \\
				& $\Bar{U} = \Bar{U}(\pi, \alpha; \bm{n})$ 		& 			& \\
\hline
\end{tabular}
\caption{Supersymmetric boundary conditions, Witten index and spontaneous supersymmetry breaking.}
\label{tab:1}
\end{center}
\end{table}

Now we are in a position to construct $\mathscr{N} = 2$ supercharges.
Let us first work in the basis in which the unitary matrix $Z$ becomes diagonal.
In this basis both the Hamiltonian $H$ and the fermion parity $(-1)^{F} = \mathcal{Z}$ become diagonal, $H = \mathrm{diag}(A^{-}_{\alpha}A^{+}_{\alpha} - 1/L(\alpha)^{2}, A^{+}_{\alpha}A^{-}_{\alpha} - 1/L(\alpha)^{2})$ and $\mathcal{Z} = \mathrm{diag}(+1, -1)$, and the wavefunction $\bm{\psi} \in \mathcal{H}$ becomes $\bm{\psi} = (\psi_{+}, \psi_{-})^{T}$.
Nilpotent supercharges are then defined as the following standard forms:
\begin{align}
Q^{+}
= 	\begin{pmatrix}
	0 			& 0 \\
	A^{+}_{\alpha} 	& 0
	\end{pmatrix}
\quad\text{and}\quad
Q^{-}
= 	\begin{pmatrix}
	0 	& A^{-}_{\alpha} \\
	0 	& 0
	\end{pmatrix}. \label{eq:2.26}
\end{align}
It is easy to see that the set of operators $\{H, \mathcal{Z}, Q^{+}, Q^{-}\}$ satisfy the $\mathscr{N}=2$ supersymmetry algebra with the trivial central extension
\begin{subequations}
\begin{align}
&\mathcal{Z}^{2} = 1, \label{eq:2.27a}\\
&(Q^{\pm})^{2} = 0, \label{eq:2.27b}\\
&Q^{\pm}\mathcal{Z} = -\mathcal{Z}Q^{\pm}, \label{eq:2.27c}\\
&Q^{+}Q^{-} + Q^{-}Q^{+} = H + \frac{1}{L(\alpha)^{2}}. \label{eq:2.27d}
\end{align}
\end{subequations}
Notice that, in terms of the self-adjoint supercharges $Q_{1} = Q^{+} + Q^{-}$ and $Q_{2} = -i(Q^{+} - Q^{-})$, Eq.~\eqref{eq:2.27d} becomes the standard anticommutation relation $\{Q_{\alpha}, Q_{\beta}\} = 2\delta_{\alpha\beta}(H + 1/L(\alpha)^{2})$.
We also note that the nilpotent supercharges can be defined in the following basis independent way:
\begin{align}
Q^{+}
= 	A^{+}_{\alpha} \otimes XP_{+}
\quad\text{and}\quad
Q^{-}
= 	A^{-}_{\alpha} \otimes XP_{-}, \label{eq:2.28}
\end{align}
whose interpretations are clear: When acting on the wavefunction $\bm{\psi} = \psi_{+}\bm{e}_{+} + \psi_{-}\bm{e}_{-} \in \mathcal{H}$, the projection operators $P_{\pm}$ pick up the modes $\psi_{\pm}\bm{e}_{\pm}$.
Then $X$ shifts $\bm{e}_{\pm}$ to $\bm{e}_{\mp}$ and $A^{\pm}_{\alpha}$ shift $\psi_{\pm}$ to $\psi_{\mp}$, yielding the maps $\psi_{\pm}\bm{e}_{\pm} \stackrel{Q^{\pm}}{\mapsto} \psi_{\mp}\bm{e}_{\mp}$.
It should be pointed out that the trivial center $1/L(\alpha)^{2} = \tan^{2}(\alpha/2)/L_{0}^{2}$ vanishes at $\alpha = 0 \pmod{2\pi}$, which corresponds to the case studied in Ref.~\cite{Nagasawa:2002un}.

\paragraph{Spectral duality web.}
It is worth pointing out here that the supersymmetric boundary conditions are mutually related by discrete transformations.
To see this, let $\mathcal{R}$ be a $\mathbb{Z}_{2}$ transformation defined by $\mathcal{R}: \bm{\psi}(x) \mapsto (\mathcal{R}\bm{\psi})(x) := \bm{\psi}(\ell - x)$, which is a reflection around the point $x = \ell/2$.
As discussed in \cite{Ohya:2012qz}, $\mathcal{R}$ is a spectrum-preserving transformation and induces the maps $U \stackrel{\mathcal{R}}{\mapsto} \Bar{U}$ and $\Bar{U} \stackrel{\mathcal{R}}{\mapsto} U$.
It is easy to see that the unitary matrices $\{U, \Bar{U}\}$ for type (D, D) and (R, R) listed in Table \ref{tab:1} are transformed into each other by the combined transformation $\mathcal{RX}(= \mathcal{XR})$ such that type (D, D) and type (R, R) are exactly isospectral.
On the other hand, the unitary matrices $\{U, \Bar{U}\}$ for type (D, R) and (R, D) are transformed into each other by the combined transformation $\mathcal{RS}(= \mathcal{SR})$, where $\mathcal{S}$ is the signature change transformation given in \cite{Cheon:2000tq} and defined by $U(\alpha_{+}, \alpha_{-}; \bm{n}) \stackrel{\mathcal{S}}{\mapsto} U(-\alpha_{+}, -\alpha_{-}; \bm{n})$ and  $\Bar{U}(\Bar{\alpha}_{+}, \Bar{\alpha}_{-}; \bm{n}) \stackrel{\mathcal{S}}{\mapsto} \Bar{U}(-\Bar{\alpha}_{+}, -\Bar{\alpha}_{-}; \bm{n})$, which do not preserve the energy spectrum.\footnote{In Ref.~\cite{Ohya:2012qz} we missed $\mathcal{S}$, which led to the wrong statement that type (D, R) and (R, D) were $\mathcal{R}$-dual and isospectral.}
However, $\mathcal{S}$ preserves the parameter dependences of energy spectrum because it just flips the eigenphases.
Hence one may say that both type (D, R) and type (R, D) belong to the same spectral family.
To summarize, we get the following web of spectral dualities:
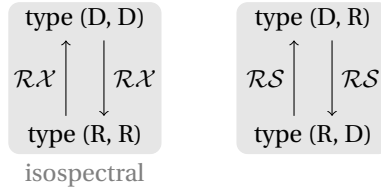
\begin{center}
\begin{tikzpicture}
\tikzstyle{every node}=[font=\small]
\begin{scope}
\filldraw[rounded corners,black!10] (-1,-0.5) rectangle (1,1.5);
\draw[->] (-0.25,0) -- node[left]{$\mathcal{RX}$} (-0.25,1);
\draw[<-] (0.25,0) -- node[right]{$\mathcal{RX}$} (0.25,1);
\draw (0,0) node[below]{type (R, R)};
\draw (0,1) node[above]{type (D, D)};
\draw[black!50] (0,-0.5) node[below]{isospectral};
\end{scope}
\begin{scope} [xshift=3cm]
\filldraw[rounded corners,black!10] (-1,-0.5) rectangle (1,1.5);
\draw[->] (-0.25,0) -- node[left]{$\mathcal{RS}$} (-0.25,1);
\draw[<-] (0.25,0) -- node[right]{$\mathcal{RS}$} (0.25,1);
\draw (0,0) node[below]{type (R, D)};
\draw (0,1) node[above]{type (D, R)};
\end{scope}
\end{tikzpicture}
\end{center}

\subsection{Superspectrum} \label{sec:2.3}
In order to compute Berry's connection explicitly, we have to find the normalized energy eigenfunctions.
In this section we solve the Schr\"{o}dinger equation for the type (D, D) and (D, R) boundary conditions and derive the superspectrum.
As mentioned above, type (R, R) is exactly isospectral to type (D, D), and energy eigenvalues of type (R, D) and type (D, R) are just related by the replacement $\alpha \mapsto -\alpha \pmod{2\pi}$.
We will see that $\mathscr{N} = 2$ supersymmetry is unbroken in the type (D, D) boundary conditions and spontaneously broken in the type (D, R) boundary conditions; see Table \ref{tab:1}.

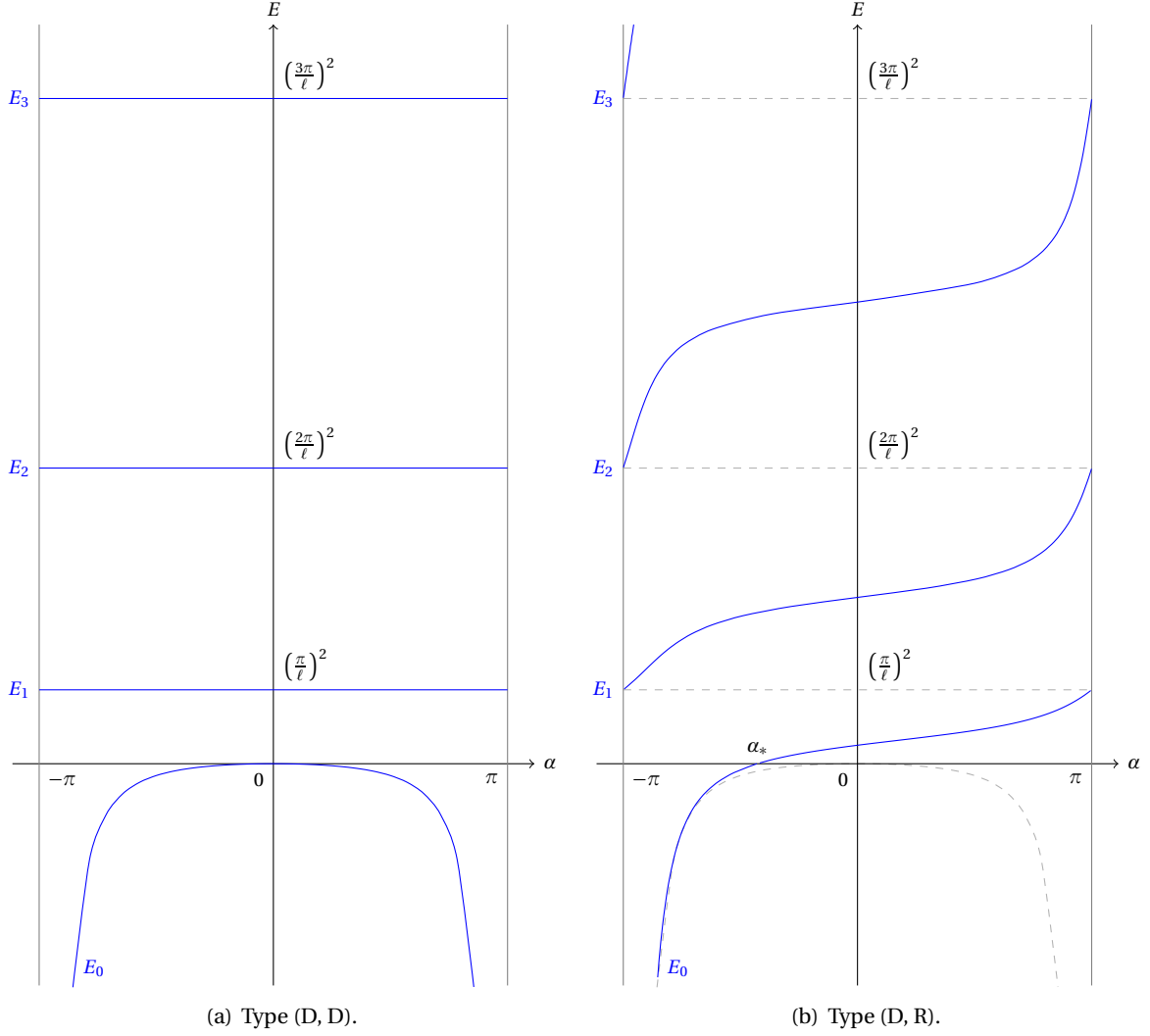
\begin{figure}[t]
\begin{center}
%
%
\subfigure[Type (D, D).]{
\begin{tikzpicture}
\tikzstyle{every node}=[font=\scriptsize]
\def\XMAX{3.5}
\def\YMAX{10}
\def\YMIN{-3}
\def\Lo{2.5} 
\def\Xo{2*rad(atan(sqrt(-\YMIN)*\Lo))}
\def\b{2*rad(atan(-sqrt(-\YMIN)*\Lo*cosh(sqrt(-\YMIN)*pi)/sinh(sqrt(-\YMIN)*pi)))}
\draw[->] (-\XMAX,0) -- (\XMAX,0) node[right]{$\alpha$};
\draw[->] (0,\YMIN) -- (0,\YMAX) node[above]{$E$};
\draw[black!50] (-pi,\YMIN) -- (-pi,\YMAX);
\draw[black!50] (pi,\YMIN) -- (pi,\YMAX);
\draw[blue,smooth] plot[domain=-\Xo:\Xo] (\x,{-tan(deg(\x/2))*tan(deg(\x/2))/(\Lo*\Lo)});
\draw[blue] (-pi,1) -- (pi,1);
\draw[blue] (-pi,4) -- (pi,4);
\draw[blue] (-pi,9) -- (pi,9);
\draw (0,0) node[below left]{$0$};
\draw (-pi,0) node[below right]{$-\pi$};
\draw (pi,0) node[below left]{$\pi$};
\draw (0,1) node[above right]{$\left(\frac{\pi}{\ell}\right)^{2}$};
\draw (0,4) node[above right]{$\left(\frac{2\pi}{\ell}\right)^{2}$};
\draw (0,9) node[above right]{$\left(\frac{3\pi}{\ell}\right)^{2}$};
\draw[blue] ({-\Xo},\YMIN) node[above right]{$E_{0}$};
\draw[blue] (-pi,1) node[left]{$E_{1}$};
\draw[blue] (-pi,4) node[left]{$E_{2}$};
\draw[blue] (-pi,9) node[left]{$E_{3}$};
\end{tikzpicture}
\label{fig:2a}}
%
%
\subfigure[Type (D, R).]{
\begin{tikzpicture}
\tikzstyle{every node}=[font=\scriptsize]
\def\XMAX{3.5}
\def\YMAX{10}
\def\YMIN{-3}
\def\Lo{2.5} 
\def\Xo{2*rad(atan(sqrt(-\YMIN)*\Lo))}
\def\b{2*rad(atan(-sqrt(-\YMIN)*\Lo*cosh(sqrt(-\YMIN)*pi)/sinh(sqrt(-\YMIN)*pi)))}
\draw[black!30,smooth,dashed] plot[domain=-\Xo:\Xo] (\x,{-tan(deg(\x/2))*tan(deg(\x/2))/(\Lo*\Lo)});
\draw[black!30,dashed] (-pi,1) -- (pi,1);
\draw[black!30,dashed] (-pi,4) -- (pi,4);
\draw[black!30,dashed] (-pi,9) -- (pi,9);
\draw[->] (-\XMAX,0) -- (\XMAX,0) node[right]{$\alpha$};
\draw[->] (0,\YMIN) -- (0,\YMAX) node[above]{$E$};
\draw[black!50] (-pi,\YMIN) -- (-pi,\YMAX);
\draw[black!50] (pi,\YMIN) -- (pi,\YMAX);
\draw[blue,smooth] plot[domain=\b-0.2:0] ({2*rad(atan(-sqrt(-\x)*\Lo*cosh(sqrt(-\x)*pi)/sinh(sqrt(-\x)*pi)))},\x);
\draw[blue,smooth] plot[domain=0.01:0.99] ({2*rad(atan(-sqrt(\x)*\Lo*cot(deg(sqrt(\x)*pi))))},\x);
\draw[blue,smooth] plot[domain=1.01:3.99] ({2*rad(atan(-sqrt(\x)*\Lo*cot(deg(sqrt(\x)*pi))))},\x);
\draw[blue,smooth] plot[domain=4.01:8.99] ({2*rad(atan(-sqrt(\x)*\Lo*cot(deg(sqrt(\x)*pi))))},\x);
\draw[blue,smooth] plot[domain=9.01:\YMAX] ({2*rad(atan(-sqrt(\x)*\Lo*cot(deg(sqrt(\x)*pi))))},\x);
\draw (0,0) node[below left]{$0$};
\draw (-pi,0) node[below right]{$-\pi$};
\draw (pi,0) node[below left]{$\pi$};
\draw ({-2*rad(atan(\Lo/pi))},0) node[above]{$\alpha_{\ast}$};
\draw (0,1) node[above right]{$\left(\frac{\pi}{\ell}\right)^{2}$};
\draw (0,4) node[above right]{$\left(\frac{2\pi}{\ell}\right)^{2}$};
\draw (0,9) node[above right]{$\left(\frac{3\pi}{\ell}\right)^{2}$};
\draw[blue] ({-\Xo},\YMIN) node[above right]{$E_{0}$};
\draw[blue] (-pi,1) node[left]{$E_{1}$};
\draw[blue] (-pi,4) node[left]{$E_{2}$};
\draw[blue] (-pi,9) node[left]{$E_{3}$};
\end{tikzpicture}
\label{fig:2b}}
\caption{
$\alpha$-dependence of energy eigenvalues $\{E_{0}, E_{1}, E_{2}, \cdots\}$.
Spectrum has the period $2\pi$.
(a) The energy eigenvalues are independent of $\alpha$ except for the ground state energy $E_{0} = -1/L(\alpha)^{2}$.
(b) The ground state energy $E_{0} > -1/L(\alpha)^{2}$ crosses zero at $\alpha = \alpha_{\ast} := -2\arctan(L_{0}/\ell)$. Notice that the energy eigenvalue $E_{n} = E_{n}(\alpha)$ satisfies $E_{n}(\alpha + 2\pi) = E_{n+1}(\alpha)$ for any $n$.}
\label{fig:2}
\end{center}
\end{figure}

\paragraph{Unbroken phase.}
Let us first consider the type (D, D) boundary conditions \eqref{eq:2.24a}.
In this case normalized energy eigenfunctions take the following forms:
\begin{subequations}
\begin{align}
\bm{\psi}_{+,n}(x)
&= 	\sqrt{\frac{2}{\ell}}\sin\left(\sqrt{E_{n}}x\right)\bm{e}_{+}, \label{eq:2.29a}\\
\bm{\psi}_{-,n}(x)
&= 	\sqrt{\frac{2}{\ell}\frac{1}{1 + E_{n}L(\alpha)^{2}}}
	\left[\sin\left(\sqrt{E_{n}}x\right)
	+ \sqrt{E_{n}}L(\alpha)\cos\left(\sqrt{E_{n}}x\right)\right]\bm{e}_{-}, \label{eq:2.29b}
\end{align}
\end{subequations}
where the energy eigenvalues $\{E_{n}\}$ are given by
\begin{align}
E_{n} = \left(\frac{n\pi}{\ell}\right)^{2} \quad (n=1,2,3,\cdots). \label{eq:2.30}
\end{align}
In addition to the above positive energy states, there is a single negative energy eigenstate $\bm{\psi}_{-,0}$ in the ``fermionic'' sector $\mathcal{H}_{-}$ which is the zero-mode of $Q^{-}$ and the ground state of the model.
A normalized eigenfunction has the following exponential form:
\begin{align}
\bm{\psi}_{-,0}(x)
&= 	\sqrt{\frac{2}{L(\alpha)}\frac{1}{\mathrm{e}^{2\ell/L(\alpha)} - 1}}
	\exp\left(\frac{x}{L(\alpha)}\right)\bm{e}_{-}, \label{eq:2.31}
\end{align}
whose energy eigenvalue (ground state energy) is given by
\begin{align}
E_{0} = -\frac{1}{L(\alpha)^{2}}. \label{eq:2.32}
\end{align}
The $\alpha$-dependence of energy spectrum is depicted in Figure \ref{fig:2a}.
In this case the ground state $\bm{\psi}_{-,0}$ is annihilated by the supercharges such that the supersymmetry is unbroken.
Indeed, the Witten index \cite{Witten:1982df} $\mathrm{Tr}\,(-1)^{F} = \mathrm{Tr}\,((-1)^{F}\mathrm{e}^{-\beta(H + 1/L(\alpha)^{2})})$, which is equivalent to the analytical index $\mathrm{ind}\,Q_{1} = \dim\ker Q^{+} - \dim\ker Q^{-} = \#(\text{zero-mode of $Q^{+}$}) - \#(\text{zero-mode of $Q^{-}$})$, is non-vanishing:
\begin{align}
\mathrm{Tr}\,(-1)^{F} = 0 - 1 = -1. \label{eq:2.33}
\end{align}

\paragraph{Broken phase.}
Let us next consider the type (D, R) boundary conditions \eqref{eq:2.24c}.
Normalized energy eigenfunctions are turned out to be of the following forms:
\begin{subequations}
\begin{align}
\bm{\psi}_{+,n}(x)
&= 	\sqrt{\frac{2}{\ell + \frac{L(\alpha)}{1 + E_{n}L(\alpha)^{2}}}}
	\sin\left(\sqrt{E_{n}}x\right)\bm{e}_{+}, \label{eq:2.34a}\\
\bm{\psi}_{-,n}(x)
&= 	\sqrt{\frac{2}{\ell + \frac{L(\alpha)}{1 + E_{n}L(\alpha)^{2}}}}
	\sin\left(\sqrt{E_{n}}(\ell - x)\right)\bm{e}_{-}, \label{eq:2.34b}
\end{align}
\end{subequations}
where $E_{n}$ ($n=0,1,2,\cdots$) are given by the real roots of the transcendental equation
\begin{align}
\tan\left(\sqrt{E}\ell\right) = -\sqrt{E}L(\alpha). \label{eq:2.35}
\end{align}
Notice that all the energy eigenvalues depend on the value of $\alpha$, and the $\alpha$-dependence is given by the inverse function of $\alpha(E) = -2\arctan(\sqrt{E}L_{0}\cot(\sqrt{E}\ell))$, which is depicted in Figure \ref{fig:2b}.
Note also that, for $-\ell < L(\alpha) < 0$ (or $-\pi < \alpha < \alpha_{\ast} := -2\arctan(L_{0}/\ell)$), the ground state energy $E_{0}$ becomes negative.
As opposed to the previous case, in the type (D, R) boundary conditions the ground states are doubly degenerate, both of which are not annihilated by the supercharges such that $\mathscr{N} = 2$ supersymmetry is spontaneously broken.
In this case the Witten index vanishes:
\begin{align}
\mathrm{Tr}\,(-1)^{F} = 0 - 0 = 0. \label{eq:2.36}
\end{align}

\section{Non-Abelian geometric phase} \label{sec:3}
So far we have constructed a class of $\mathscr{N} = 2$ mechanics which, except for the ground states of type (D, D) and (R, R), enjoys doubly-degenerate energy spectrum.
The parameter space of these models is $\mathcal{M}_{\text{SUSY}} = U(1) \times U(2)/U(1)^{2} \ni (\alpha, \bm{n})$, where the first $U(1)$ factor controls the energy eigenvalues and the second $U(2)/U(1)^{2}$ factor defines isospectral family.
In this section we consider a time-dependent situation in which the $U(1)$ parameter is kept fixed\footnote{Varying the $U(1)$ parameter gives rise to a different story. Since $E_{n}(\alpha + 2\pi) = E_{n+1}(\alpha)$ for the type (D, R) boundary conditions (see Figure \ref{fig:2b}), if one varies the parameter $\alpha$ and completes a cycle along $U(1) \cong S^{1}$, one gets $\bm{\psi}_{\pm,n}(x; \alpha + 2\pi) = \bm{\psi}_{\pm,n+1}(x; \alpha)$; that is, the final state does not belong to initial state's eigenspace, which is called Cheon's anholonomy \cite{Cheon:1998pt}. Though Cheon's anholonomy is an interesting subject itself, in this note we do not touch upon the question of adiabatic time-evolution along a closed loop on the whole supersymmetric parameter space $\mathcal{M}_{\text{SUSY}}$.} yet the $U(2)/U(1)^{2}$ parameters are externally driven and a state evolves according to the time-dependent formulation of quantum mechanics.
For a time-evolution from time $t = 0$ to time $t = T$, transitions between different eigenspaces $\mathcal{H}_{n} = \mathrm{span}\{|\bm{\psi}_{+,n}(\bm{n}(t))\rangle, |\bm{\psi}_{-,n}(\bm{n}(t))\rangle\}$ and $\mathcal{H}_{m} = \mathrm{span}\{|\bm{\psi}_{+,m}(\bm{n}(t))\rangle, |\bm{\psi}_{-,m}(\bm{n}(t))\rangle\}$ ($n \neq m$) are suppressed by the factor $1/T$ such that different eigenspaces are decoupled in the adiabatic limit $T \to \infty$.
Namely, if an initial state $|\bm{\psi}_{\text{in}}(\bm{n}(0))\rangle = c_{+,n}(0)|\bm{\psi}_{+,n}(\bm{n}(0))\rangle + c_{-,n}(0)|\bm{\psi}_{-,n}(\bm{n}(0))\rangle$ is an element of the subspace $\mathcal{H}_{n}$, the final state $|\bm{\psi}_{\text{out}}(\bm{n}(T))\rangle = c_{+,n}(T)|\bm{\psi}_{+,n}(\bm{n}(T))\rangle + c_{-,n}(T)|\bm{\psi}_{-,n}(\bm{n}(T))\rangle$ remains in the subspace $\mathcal{H}_{n}$ and the time-evolution is given by the $2 \times 2$ unitary matrix, $\left(\begin{smallmatrix}c_{+,n}(T)\\ c_{-,n}(T)\end{smallmatrix}\right) = U_{T}\left(\begin{smallmatrix}c_{+,n}(0)\\ c_{-,n}(0)\end{smallmatrix}\right)$.
Under an adiabatic time-evolution along a closed path $\gamma: [0, T] \to U(2)/U(1)^{2}$ with $\gamma(T) = \gamma(0)$, the unitary matrix $U_{T}$ consists of a $T$-dependent trivial dynamical phase $\exp(-iE_{n}T)$ as well as a $T$-independent nontrivial non-Abelian geometric phase $\Gamma_{A}(\gamma)$ given by the path-ordered exponential \cite{Wilczek:1984dh}
\begin{align}
\Gamma_{A}(\gamma)
&= 	\mathcal{P}\exp\left(i\oint_{\gamma}A\right). \label{eq:3.1}
\end{align}
Here $A = \left(\begin{smallmatrix}A_{++}&A_{+-}\\A_{-+}&A_{--}\end{smallmatrix}\right)$ is a $2 \times 2$ hermitian matrix-valued 1-form (Berry's connection) on $U(2)/U(1)^{2} \cong S^{2}$ given by
\begin{align}
A_{\alpha\beta}
&= 	i\langle\bm{\psi}_{\alpha, n}|d|\bm{\psi}_{\beta, n}\rangle
:= 	i\int_{0}^{\ell}\!\!dx\,\bm{\psi}_{\alpha, n}^{\dagger}(x)d\bm{\psi}_{\beta, n}(x), \quad
	(\alpha, \beta = \pm), \label{eq:3.2}
\end{align}
where $d$ stands for the exterior derivative on the parameter space $U(2)/U(1)^{2} \cong S^{2}$.
The field strength 2-form is given by
\begin{align}
F
&= 	dA - iA \wedge A. \label{eq:3.3}
\end{align}
Under the unitary change of basis $\bm{\psi}_{\alpha,n} \mapsto \Tilde{\bm{\psi}}_{\alpha,n} = \bm{\psi}_{\beta,n} g_{\beta\alpha}$ ($g \in U(2)$), the Berry connection \eqref{eq:3.2} and the field strength 2-form \eqref{eq:3.3} transform as follows:
\begin{subequations}
\begin{align}
A
&\mapsto \Tilde{A} = g^{\dagger}Ag + ig^{\dagger}dg, \label{eq:3.4a}\\
F
&\mapsto \Tilde{F} = g^{\dagger}Fg. \label{eq:3.4b}
\end{align}
\end{subequations}
Accordingly, the non-Abelian geometric phase changes as
\begin{align}
\Gamma_{A}(\gamma)
&\mapsto \Gamma_{\Tilde{A}}(\gamma)
= 	g^{\dagger}\Gamma_{A}(\gamma)g. \label{eq:3.5}
\end{align}
The goal of this section is to compute $A$ and $F$ in two different gauges (``string'' gauge and ``hedgehog'' gauge) and to show that the Berry connection is given by the Wu-Yang-like magnetic monopole first discovered by Moody, Shapere and Wilczek in the context of adiabatic decoupling limit of diatomic molecule\cite{Moody:1985ty}.

\subsection{\texorpdfstring{$SU(2)$}{SU(2)} magnetic monopole} \label{sec:3.1}
Upon substituting the solutions \eqref{eq:2.29a} and \eqref{eq:2.29b} or \eqref{eq:2.34a} and \eqref{eq:2.34b} to the definition \eqref{eq:3.2}, one readily finds the Berry connection to be of the form
\begin{align}
A
= 	\begin{pmatrix}
	i\bm{e}_{+}^{\dagger}d\bm{e}_{+} 			& K_{n}(\alpha)i\bm{e}_{+}^{\dagger}d\bm{e}_{-} \\
	K_{n}(\alpha)i\bm{e}_{-}^{\dagger}d\bm{e}_{+} 	& i\bm{e}_{-}^{\dagger}d\bm{e}_{-}
	\end{pmatrix}, \label{eq:3.6}
\end{align}
where $K_{n}(\alpha) \in \mathbb{R}$ is the overlapping integral between the components $\psi_{+,n}$ and $\psi_{-,n}$ given by
\begin{align}
K_{n}(\alpha)
:= 	\int_{0}^{\ell}\!\!\!dx\,\psi_{+,n}(x; \alpha)\psi_{-,n}(x; \alpha). \label{eq:3.7}
\end{align}
Notice that, just as we did in section \ref{sec:2.3}, the components $\psi_{\pm,n}$ are always taken to be real thanks to the time-reversal invariance of both the Schr\"{o}dinger equation $h\psi_{\pm} = E\psi_{\pm}$ and the boundary conditions \eqref{eq:2.24a}--\eqref{eq:2.24d}.
The integral \eqref{eq:3.7} is exactly calculable of course, but its explicit form is not necessary for the following discussions.
The field strength 2-form $F = dA - iA \wedge A$ is easily calculated with the result
\begin{align}
F
= 	i\left(1 - K_{n}(\alpha)^{2}\right)\sigma_{3}
	\bm{e}_{-}^{\dagger}d\bm{e}_{+} \wedge \bm{e}_{+}^{\dagger}d\bm{e}_{-}. \label{eq:3.8}
\end{align}
Though not so obvious, this is nothing but the $SU(2)$ magnetic monopole in Ref.~\cite{Moody:1985ty}.
To see this, let us parameterize the unit 3-vector $\bm{n}$ into the spherical coordinates $\bm{n} = (\sin\theta\cos\phi, \sin\theta\sin\phi, \cos\theta)$, where $\theta \in [0, \pi]$ and $\phi \in [0, 2\pi)$ are the polar and azimuthal angles, respectively.
Then the orthonormal eigenvectors $\{\bm{e}_{+}, \bm{e}_{-}\}$ of the unitary matrix $Z = \bm{n} \cdot \bm{\sigma}$ are taken to be of the forms
\begin{align}
\bm{e}_{+}
= 	\begin{pmatrix}
	\cos\frac{\theta}{2} \\
	\mathrm{e}^{i\phi}\sin\frac{\theta}{2}
	\end{pmatrix}
	\quad\text{and}\quad
\bm{e}_{-}
= 	\begin{pmatrix}
	-\mathrm{e}^{-i\phi}\sin\frac{\theta}{2} \\
	\cos\frac{\theta}{2}
	\end{pmatrix}. \label{eq:3.9}
\end{align}
Substituting these into Eqs.~\eqref{eq:3.6} and \eqref{eq:3.8}, one finds that Berry's connection and the field strength 2-form become $A = A_{\theta}d\theta + A_{\phi}d\phi$ and $F = \frac{1}{2}F_{\theta\phi}d\theta \wedge d\phi$, where
\begin{subequations}
\begin{align}
A_{\theta}
&= 	- K_{n}(\alpha)\sin\phi\frac{\sigma_{1}}{2}
	+ K_{n}(\alpha)\cos\phi\frac{\sigma_{2}}{2}, \label{eq:3.10a}\\
A_{\phi}
&= 	- K_{n}(\alpha)\sin\theta\cos\phi\frac{\sigma_{1}}{2}
	- K_{n}(\alpha)\sin\theta\sin\phi\frac{\sigma_{2}}{2}
	- (1 - \cos\theta)\frac{\sigma_{3}}{2}, \label{eq:3.10b}
\end{align}
\end{subequations}
and
\begin{align}
F_{\theta\phi}
= 	-\left(1 - K_{n}(\alpha)^{2}\right)\sin\theta\frac{\sigma_{3}}{2}, \label{eq:3.11}
\end{align}
which are exactly the same forms as those in Ref.~\cite{Moody:1985ty}.
(Precisely speaking, Eqs.~\eqref{eq:3.10a}, \eqref{eq:3.10b} and \eqref{eq:3.11} coincide with Eqs. (11), (12) and (13) in Ref.~\cite{Moody:1985ty} under the change of variable $\phi \to \frac{\pi}{2} - \phi$.)
We note that these monopole configurations are rather ubiquitous and appear in various contexts of physics, such as nuclear quadrupole resonance \cite{Zee:1988} and antiferromegnets with spontaneously broken $SU(2)$ spin symmetry \cite{Chandrasekharan:2006wn}.

It should be pointed out that, since $1-\cos\theta$ does not vanish at $\theta = \pi$, the 1-form $A_{\phi}d\phi$ is ill-defined at the south pole.
In other words, Berry's connection $A$ suffers from the Dirac string singularity along the negative 3-axis.
This Dirac string is just the reflection of the fact that the basis vectors $\{\bm{e}_{+}, \bm{e}_{-}\}$ cannot be globally well-defined over the whole 2-sphere.
Indeed, the parameterization \eqref{eq:3.9} is ill-defined at the south pole because $\mathrm{e}^{\pm i\phi}\sin\frac{\theta}{2}$ do not vanish at $\theta = \pi$.
As is well-known, however, the Dirac string singularity can be removed by singular gauge transformation \cite{Arafune:1974uy}.
Before closing this note we would like to perform this and transform the Berry connection into a manifestly spherically symmetric form that is more familiar in $SU(2)$ Yang-Mills gauge theory.

\subsection{Singular gauge transformation} \label{sec:3.2}
Let us consider the gauge transformation given by the following $2 \times 2$ unitary matrix:
\begin{align}
g
&= 	\begin{pmatrix}
	\bm{e}_{+}^{\dagger} \\
	\bm{e}_{-}^{\dagger}
	\end{pmatrix}. \label{eq:3.12}
\end{align}
Notice that $g$ inherits the Dirac string from the basis $\{\bm{e}_{+}, \bm{e}_{-}\}$.
Indeed, in the parameterization \eqref{eq:3.9}, Eq.~\eqref{eq:3.12} becomes $g = \left(\begin{smallmatrix}\cos\frac{\theta}{2} & \mathrm{e}^{-i\phi}\sin\frac{\theta}{2}\\ -\mathrm{e}^{i\phi}\sin\frac{\theta}{2} & \cos\frac{\theta}{2}\end{smallmatrix}\right) = \mathrm{e}^{-i\phi\sigma_{3}/2}\mathrm{e}^{i\theta\sigma_{2}/2}\mathrm{e}^{i\phi\sigma_{3}/2}$, which is ill-defined at the south pole $\theta = \pi$ and suffers from the Dirac string singularity along the negative 3-axis.

Now, as shown in appendix \ref{appendix:A}, in this gauge the Berry connection \eqref{eq:3.4a} and the field strength 2-form \eqref{eq:3.4b} take the following simple forms:
\begin{subequations}
\begin{align}
\Tilde{A}
&= 	\frac{i}{2}\left(1 - K_{n}(\alpha)\right)ZdZ, \label{eq:3.13a}\\
\Tilde{F}
&= 	-\frac{i}{4}\left(1 - K_{n}(\alpha)^{2}\right)ZdZ \wedge ZdZ. \label{eq:3.13b}
\end{align}
\end{subequations}
Note that $\Tilde{A}$ vanishes at $K_{n}(\alpha) = 1$ and becomes pure gauge $iZdZ$ at $K_{n}(\alpha) = -1$, both of which yield the vanishing field strength $\Tilde{F} = 0$.
For the following discussions it is convenient to parameterize the unit 3-vector $\bm{n}$ in $Z = \bm{n} \cdot \bm{\sigma}$ into the following ``hedgehog'' configuration:
\begin{align}
\bm{n} = \frac{\bm{r}}{r}, \label{eq:3.14}
\end{align}
where $\bm{r} = (x_{1}, x_{2}, x_{3}) \in \mathbb{R}^{3}\setminus\{\bm{0}\}$ is a nonzero real 3-vector and $r = \sqrt{x_{1}^{2} + x_{2}^{2} + x_{3}^{2}}$ is its length.
A straightforward calculation gives $\Tilde{A} = \Tilde{A}_{i}dx_{i}$ and $\Tilde{F} = \frac{1}{2}\Tilde{F}_{ij}dx_{i} \wedge dx_{j}$, where
\begin{subequations}
\begin{align}
\Tilde{A}_{i}
&= 	\epsilon_{ijk}\frac{x_{j}}{r^{2}}\frac{\sigma_{k}}{2}\left(1 - K_{n}(\alpha)\right), \label{eq:3.15a}\\
\Tilde{F}_{ij}
&= 	-\epsilon_{ijk}\frac{x_{k}x_{l}}{r^{4}}\frac{\sigma_{l}}{2}\left(1 - K_{n}(\alpha)^{2}\right). \label{eq:3.15b}
\end{align}
\end{subequations}
Now it is obvious that, when $K_{n}(\alpha) = 0$, Berry's connection \eqref{eq:3.15a} becomes the Wu-Yang magnetic monopole which is a classical solution of pure $SU(2)$ Yang-Mills gauge theory (see e.g. Ref.~\cite{Actor:1979in} for review).
Note, however, that for nonzero $K_{n}(\alpha)$ the Berry connection \eqref{eq:3.15a} is not a solution to the sourceless Yang-Mills equations except for the trivial cases $K_{n}(\alpha) = \pm 1$.
Note also that, if $K_{n}$ was a function of $r$, the Berry connection \eqref{eq:3.15a} would be the celebrated 't Hooft-Polyakov monopole \cite{'tHooft:1974qc,Polyakov:1974ek} in Yang-Mills-Higgs theory.\footnote{The 't Hooft-Polyakov monopole has been realized as Berry's connection in the rather different context of supersymmetric quantum mechanics \cite{Pedder:2007ff,Sonner:2008be}.}

Though the Dirac string singularity disappears in this gauge, it becomes apparent that there is another singularity at $r = 0$ where both $\Tilde{A}_{i}$ and $\Tilde{F}_{ij}$ blow up.
This singularity is merely a reflection of the fact that the parameterization \eqref{eq:3.14} is ill-defined at $r = 0$.
It should be emphasized that, as opposed to the standard Berry connection \cite{Berry:1984jv}, this singularity is neither a signal of energy level crossings nor a signal of additional spectral degeneracies at $r = 0$.
(Recall that the energy eigenvalues do not depend on $Z \in U(2)/U(1)^{2} \cong S^{2}$ in any way.)

\subsection*{Acknowledgement}
The author is supported in part by ESF grant CZ.1.07/2.3.00/30.0034.

\appendix
\titleformat{\section}[block]{\filright\bfseries\mathversion{bold}}{Appendix \thesection.}{0.5em}{}[\titlerule]
\section{Computational details} \label{appendix:A}
In this section we derive Eqs.~\eqref{eq:3.13a} and \eqref{eq:3.13b}.
With the unitary matrix $g$ given in \eqref{eq:3.12}, the Berry connection \eqref{eq:3.6} is transformed according to the law \eqref{eq:3.4a}:
\begin{align}
\Tilde{A}
&= 	\begin{pmatrix}
	\bm{e}_{+} 	& \bm{e}_{-}
	\end{pmatrix}
	\begin{pmatrix}
	i\bm{e}_{+}^{\dagger}d\bm{e}_{+} 		& K_{n}i\bm{e}_{+}^{\dagger}d\bm{e}_{-} \\
	K_{n}i\bm{e}_{-}^{\dagger}d\bm{e}_{+} 	& i\bm{e}_{-}^{\dagger}d\bm{e}_{-}
	\end{pmatrix}
	\begin{pmatrix}
	\bm{e}_{+}^{\dagger} \\
	\bm{e}_{-}^{\dagger}
	\end{pmatrix}
	- i
	\begin{pmatrix}
	d\bm{e}_{+} 	& d\bm{e}_{-}
	\end{pmatrix}
	\begin{pmatrix}
	\bm{e}_{+}^{\dagger} \\
	\bm{e}_{-}^{\dagger}
	\end{pmatrix} \nonumber\\
&= 	iK_{n}\bm{e}_{+}\bm{e}_{+}^{\dagger}d\bm{e}_{-}\bm{e}_{-}^{\dagger}
	+ iK_{n}\bm{e}_{-}\bm{e}_{-}^{\dagger}d\bm{e}_{+}\bm{e}_{+}^{\dagger}
	- i(\bm{1}_{2} - \bm{e}_{-}\bm{e}_{-}^{\dagger})d\bm{e}_{-}\bm{e}_{-}^{\dagger}
	- i(\bm{1}_{2} - \bm{e}_{+}\bm{e}_{+}^{\dagger})d\bm{e}_{+}\bm{e}_{+}^{\dagger} \nonumber\\
&= 	i(K_{n} - 1)
	(\bm{e}_{+}\bm{e}_{+}^{\dagger}d\bm{e}_{-}\bm{e}_{-}^{\dagger}
	+ \bm{e}_{-}\bm{e}_{-}^{\dagger}d\bm{e}_{+}\bm{e}_{+}^{\dagger}), \label{eq:A.1}
\end{align}
where the first equality follows from the identity $g^{\dagger}dg = -dg^{\dagger}g$ and the last equality follows from the completeness relations $\bm{1}_{2} - \bm{e}_{\mp}\bm{e}_{\mp}^{\dagger} = \bm{e}_{\pm}\bm{e}_{\pm}^{\dagger}$.
Making use of the identities $P_{\pm}dP_{\pm} = -P_{\pm}dP_{\mp} = -\bm{e}_{\pm}\bm{e}_{\pm}^{\dagger}(d\bm{e}_{\mp}\bm{e}_{\mp}^{\dagger} + \bm{e}_{\mp}d\bm{e}_{\mp}^{\dagger}) = -\bm{e}_{\pm}\bm{e}_{\pm}^{\dagger}d\bm{e}_{\mp}\bm{e}_{\mp}^{\dagger}$, where $P_{\pm} = \bm{e}_{\pm}\bm{e}_{\pm}^{\dagger}$ are the projection operators satisfying $dP_{+} + dP_{-} = 0$, we get
\begin{align}
\Tilde{A}
&= 	i(1 - K_{n})(P_{+}dP_{+} + P_{-}dP_{-}). \label{eq:A.2}
\end{align}
It follows from the alternative equivalent expressions for the projection operators $P_{\pm} = (\bm{1}_{2} \pm Z)/2$ that Eq.~\eqref{eq:A.2} reduces to the following compact form:
\begin{align}
\Tilde{A}
&= 	\frac{i}{2}(1 - K_{n})ZdZ. \label{eq:A.3}
\end{align}
The field strength 2-form is therefore
\begin{align}
\Tilde{F}
&= 	d\Tilde{A} - i\Tilde{A} \wedge \Tilde{A}
= 	\frac{i}{2}(1 - K_{n})dZ \wedge dZ + \frac{i}{4}(1 - K_{n})^{2}ZdZ \wedge ZdZ \nonumber\\
&= 	-\frac{i}{4}(1 - K_{n}^{2})ZdZ \wedge ZdZ, \label{eq:A.4}
\end{align}
where in the last line we have used $ZdZ \wedge ZdZ = -dZ \wedge dZ$, which follows from the identities $ZdZ = -dZZ$ and $Z^{2} = \bm{1}_{2}$.

\bibliographystyle{utphys}
\bibliography{Bibliography}

\end{document}